# Space Diplomacy and the Artemis Accords

*Nancy Riordan* | ORCID: 0000-0002-2403-7268
Doctoral candidate in Global Governance and Human Security, John W. McCormack Graduate School of Policy and Global Studies, University of Massachusetts, Boston, MA, United States
*nancy.riordan@umb.edu*

*Miloslav Machoň* | ORCID: 0000-0001-8591-3770
Assistant Professor and Research Fellow, Department of International and Diplomatic Studies, Prague University of Economics and Business, Prague, Czech Republic
*miloslav.machon@vse.cz*

*Lucia Csajková* | ORCID: 0000-0003-3015-7167
Research Fellow, Department of International and Diplomatic Studies Department, Prague University of Economics and Business, Prague, Czech Republic
*csal00@vse.cz*

**Summary**

The growth of human activity in outer space is attracting more International Relations (IR) scholar's attention, enabling an understanding of the involvement of specific groups of actors and the dynamics of political negotiations that lead to concluding agreements on using outer space for peaceful purposes. This paper provides analysis based on the triangulation of qualitative data gathered via document analysis and in-depth semi-structured expert interviews to gain insight into the involvement of the actors responsible for the negotiations that led to the Artemis Accords and their diplomatic communication style. The results identified different uses of public and private diplomatic communication for advancing norms of behaviour and transparency. Negotiators used public diplomatic communication in order to influence foreign governments on the need for norms of behaviour and transparency to further peaceful space exploration beyond low Earth orbit. Private diplomatic communication facilitated the inclusion of commercial partners.

## 1    Introduction[1]

International co-operation in science and technology domains and its relation to foreign policy have attracted the attention of International Relations (IR) scholars across decades.[2] Such scholars have used diverse quantitative and qualitative methods to provide findings about the political implications for scientific and technological co-operative projects.[3] In the space policy domain, emerging research has begun to consider the mosaic of relations among a wide range of governmental and non-governmental actors and their interplay with epistemic communities, especially in the context of space sustainability governance.[4] Previous research published in *The Hague Journal of Diplomacy* has focused on joint international operations in outer space, in the context of public diplomacy. However, space-related activities constituted only particular topics, such as those related to UN Security Council membership and diplomatic gift-giving.[5]

---

[1] First, we want to thank our eight interviewees for their readiness to provide detailed insight about the negotiation process for the Artemis Accords. We also profoundly acknowledge three anonymous reviewers, Professors Mai'a K. Davis Cross and Saadia M. Pekkanen, and the *HJD* editors for guiding us and providing constructive feedback.

[2] Skolnikoff 2001, 462–468.

Now is a critical time to take stock of where and how space diplomacy is put into practice.[6] Whether it is the increase in the number of countries establishing space forces and still others seeking space-enabled military capabilities, tensions over dual-use capabilities and anti-satellite weapons tests, or ongoing escalations involving a variety of stakeholders over non-geostationary orbit mega-constellations, diplomatic efforts of all types are needed to keep the peace in space.[7] As Paul Meyer wrote in 2018, a 'revival of diplomatic activism on behalf of space security is required, not only by concerned states, but also on the part of the wider stakeholder community including the private sector and civil society that benefits from the current regime'.[8] Moreover, ambassador W. Robert Pearson and astrophysicist Dr Benjamin L. Schmitt called for action in their article 'The Crisis in Space':

> [n]o single government or company can define the future of space. International diplomacy must be supported by practitioners of science and technology taking an active role when the next phase of space policy is crafted…before these issues lead to open conflict, diplomacy and science should be used to fashion measures regulating resource exploitation, manage military use to provide defensive deterrence while avoiding escalation, and expand the reach of space exploration to benefit humanity. If we are to keep up with the rapid advancement of space science and meet the urgent need for global agreement, diplomacy must shape the future and not wait to react to crises.[9]

In this issue, Cross and Pekkanen have conceptualised space diplomacy as 'the formal and informal interactions, agreements, and deliberations that occur among actors involved in space, whether conducted by professional diplomats or non-state actors'.[10] The growth of activity in outer space and the need for maintaining co-operation among space actors based on equality and mutual respect to sustain the peaceful status of outer space, is attracting more scholarship, enabling an understanding of the involvement of specific groups of actors and the dynamics of political negotiations that lead to concluding agreements on using outer space for peaceful purposes.[11] Joining this increased interest in space diplomacy, this paper uncovers a backstory of endeavours in diplomatic negotiations that focuses on mechanisms of diplomatic communication between the actors involved, which led to the Artemis Accords, including a supporting role for the US National Space Council (NSpC) and its Users Advisory Group (NSpC UAG). The Accords are an international agreement meant to support the expansion of NASA's Artemis programme and its mission to establish the first long-term presence on the Moon.[12] The Artemis Accords are principles founded on the 1967 Treaty on Principles Governing the Activities of States in the Exploration and Use of Outer Space, including the Moon and Other Celestial Bodies (the Outer Space Treaty) that seek to create a safe and transparent space environment to facilitate exploration, science, and commercial activities.[13] The negotiations for the Accords are an example of diplomatic activism that did not wait around for a crisis in space and were expertly crafted during the historic Covid-19 pandemic.

---

[3] Freeman 1991, 510–511.

[4] Martinez et al. 2019, 29–31; Oltrogge and Christensen 2020, 433–435; Trur 2021, 452–453.

[5] Sievers 2021, 113; Krasnyak 2020, 399–400; Thorhallsson and Eggertsdóttir 2021, 60–61.

[6] Recent commentary during a United Nations Institute for Disarmament Research web conference #OS21: The UNIDIR Outer Space Security Conference 2021 – Several speakers suggested the time was not ripe for a new treaty-based agreement (Maceda, Krabill and Ortega 2021, 15).

[7] Russia and China established space forces in 2015; the US and others followed suit in 2018, with Japan and France enhancing their space divisions in 2019 (Aerospace Security Project at the Centre for Strategic and International Studies 2017–2021). Russia and China voted against the establishment of an UN Open-Ended Working Group then (Weeden and Samson 2022, 01-01 – 03-23; CSIS 2022; UNGA 2021, 1).

[8] Meyer 2018, 5.

[9] Pearson and Schmitt 2021.

[10] Cross and Pekkanen 2023.

[11] Di Pippo, Woltran and Stasko 2020, 27–28.

[12] NASA 2022.

[13] NASA 2022.

This paper operationalises the concept of diplomatic communication to analyse information derived from expert interviews. The analysis aims to contribute to a deeper understanding of how practitioners of science and technology as well as diplomats under the Trump presidency shaped the implementation of space policy and international engagement, and in particular supported the decision to negotiate the Artemis Accords. The article will therefore include details about the distinction between public and private communication strategies, aspects of trust in arguing, framing and rationale for drafting the agreement, interagency co-operation and expertise. It also describes the preliminary steps of the negotiation process. On the domestic and international levels, trust and the art of persuasion facilitated the negotiations. Following the literature on the concept of science diplomacy, this paper recognises the soft power of space affairs in diplomatic negotiations.[14] Therefore, it assumes that space diplomacy uses scientific knowledge and technological capabilities to reinforce the attractiveness of a country in order to influence other international actors.

In tracing the diplomatic process for the Artemis Accords, the qualitative research design used the triangulation of multiple data sources based on data gathered via eight in-depth semi-structured expert interviews (see Appendix no. 1) and document analysis. According to the principles of snowball sampling, interviewees included individuals affiliated with the US Department of State, NASA's Office of International and Interagency Relations (OIIR), the NSpC as well as practitioners and academic experts in space policy and law with specific knowledge related to the negotiations of the Accords.[15] The set of questions followed the methodological rules for systematising expert interviews with the aim of obtaining access to exclusive knowledge based on professional expertise.[16] The information gained in expert interviews and rules of systematic sampling gave rise to a set of documents for analysis. This comprised US federal policies relevant to the Artemis Accords.

## 2 Space Diplomacy and Soft Power

Public diplomacy aims to build and maintain relationships with publics in foreign countries.[17] Regarding building and maintaining relationships, Joseph S. Nye emphasised soft power as an attractive way to influence the preferences of other actors in international politics.[18] Soft power stems from cultural values embedded in foreign policy. The pull effect of soft power depends on how foreign politicians and the public perceive communicated values and whether they find the values attractive.[19] The recent growth of human activity in outer space stimulates research about conditions of co-operation among various actors in the space domain.[20] The concept of space diplomacy outlined by Cross and Pekkanen assumes a co-operative interplay between practitioners of science and technology and diplomats.[21]

Since Eugene Skolnikoff recognised the capability of arms control, high-technology competitiveness, and effective responses to global threats as science and technology factors influencing leadership in international politics,[22] science diplomacy has become a relevant concept for explaining the interplay between practitioners of science and technology and diplomats. In 2002, an issue of the journal *Science and Public Policy* confirmed the relevance of the nexus between science, technology, and foreign policy in the era of globalisation.[23] The issue labelled the USA, the EU, the Republic of Korea, and Japan as the most progressive national systems of innovation at the beginning of the 21st century. Its case studies provided empirical evidence of national perspectives on the integration of science and technology policies into foreign policy domains, and the implications presented opportunities for further generalisation.

---

[14] Nye 2004, 25.
[15] Krippendorff 2019, 121.
[16] Bogner and Menz 2009, 46–47; Petintseva, Faria and Eski 2020, 113–114.
[17] Snow 2020, 4; Melissen and Wang 2019, 1.
[18] Nye 1990, 166; Nye 2019, 8.
[19] Çevik 2019, 54.
[20] Di Pippo, Woltran and Stasko 2020, 28.
[21] Cross and Pekkanen 2023.
[22] Skolnikoff 1994, 12–13.
[23] Stein 2002, 403.

Thus, the science diplomacy concept integrates science and foreign policy domains, in which scientists and diplomats act together.[24] Dialogue, negotiation, and compromise are its foundation, and it is designed to understand, engage, inform, and influence foreign actors in support of national interests.[25] Unlike science, science diplomacy has no prerequisite for political or ideological neutrality. Following the implementation of government policy, diplomacy pursues and implements foreign policy goals in the international arena. Therefore, science diplomacy is defined as the state-led negotiation process in which the state represents itself and advocates national interests on how to acquire, utilise, and communicate scientific knowledge.[26]

## 2.1 *Space Science for Diplomacy*

Following the definitions provided by the British Royal Society and the American Association for the Advancement of Science, in the context of space diplomacy the science for diplomacy dimension might be operationalised as using scientific knowledge about outer space and knowledge about Earth collected from space for building and improving international relations.[27] Due to recent technological developments and socio-economic shifts allowing cheaper and faster access to outer space, knowledge about outer space has become more diverse. This knowledge emerges from complex interactions among an increased number and diversity of actors.[28] These include governments, private companies, academics and citizens.[29] Through diplomatic communication, actors are involved in the uses and exploration of outer space. Diplomats communicate the results of space science to support foreign policy objectives, diplomatic activities and negotiations in the space domain.

Science for diplomacy is based on the potential of space science to influence other countries. In terms of soft power, deriving the values of space diplomacy from scientific knowledge supports intelligibility and attractiveness of the values.[30] Knowledge about outer space constitutes an unprejudiced result of processing and acquiring information from empirical observation. The acquiring and processing of sensory perceptions follows scientific methodology. Universal, non-political, and non-ideological values constitute the core methods.[31]

Therefore, incorporating scientific results derived from space science into negotiations between scientists and diplomats has the potential to facilitate and streamline the building of relations among like-minded actors. Justifying diplomatic negotiations via the values of space science could play a significant role in the success of diplomatic efforts involving countries whose political relationships have been difficult. In this regard, actors engaging in space diplomacy use communication as a mechanism for influencing the preferences of other actors interested in space activities. The actors of space diplomacy communicate the results of space science among representatives of internationally recognised entities.[32]

Jönsson and Hall recognise public and private communication as two types of diplomatic exchange.[33] Public communication refers mainly to public diplomacy based on initiatives by the government of one country to influence the politicians or public opinion of another country.[34] In the space domain, public diplomatic communication aims to persuade foreign publics to regard its space policy, its values and ideas favourably. According to Nye, the means of public diplomacy involves the communication of foreign policy decisions to domestic and international media. It also includes strategic communication, such as public diplomacy campaigns and the development of relationships with individuals via conferences, scholarships, seminars, and access to media by non-state actors.[35] Thus, public diplomacy is a tool for creating a positive image in the space domain among foreign publics.[36]

---

[24] Ruffini 2017, 16.
[25] Snow 2020, 8; Banks 2020, 64.
[26] Turekian et al. 2015, 5–6.
[27] The Royal Society 2010, v–vi; Hills et al. 2022, 2.
[28] Athanasopoulos 2019, 5.
[29] Bohlmann and Petrovici 2019, 2.
[30] Chalmers 2015, 4; Galluccio 2021, 37; Turekian et al. 2015, 4.
[31] Thorson and Seo 2014, 137.
[32] Jönsson 2016, 79.
[33] Jönsson and Hall 2005, 90.
[34] Jönsson and Hall 2005, 90.
[35] Nye 2008, 101–102.
[36] Rashica 2018, 78.

In contrast, private diplomatic communication is used as a means of diplomacy to obtain information from certain individuals and non-governmental groups.[37] It can be utilised in negotiations with recognised representatives, professionals, enterprises and private industry to draft foreign and global policy.[38] Private diplomatic communication, when it occurs with professionals such as scientists, is also about acquiring, utilising and communicating scientific knowledge within different state sectors.[39] This scientific engagement reflects co-operation between political actors and science and technological experts.[40] Communication with other representatives, such as experts from commercial space industries, and communication with scientific professionals constitutes another major form of private communication. Private diplomatic communication involves means such as persuasion and trust.

Persuasion has been defined as 'a micro-causal process conceptualised as argumentation and framing, most likely to involve political and coercive dynamics'.[41] The micro-casual process aims to influence or to generate belief in something through reasoning.[42] From the perspective of political psychology, diplomatic persuasion emphasises framing in argumentation and psychological factors, such as trust in diplomatic persuasion.[43] Arguing is essential at all phases of the negotiation process.[44] It is a form of mutual persuasion in which differences are argued in dialogue.[45] Argumentation should, as Kerr states, lead to 'a change in the interests or behaviour of the actors concerned'.[46] Scholars in political psychology incorporate framing in argumentation, as they perceive framing as crucial for shaping opinions on relevant policy topics.[47] Framing effects occur when equivalent representations of the same policy problem are communicated systematically with different decision goals.[48]

In political psychology, trust is also a crucial factor for understanding the outcomes of diplomatic persuasion. In this model, trustworthiness is perceived as a personal attribute of the message-sender.[49] Trust during scientific engagements requires, as Thorson and Seo describe, 'longer-term involvement of the parties, since it takes time and demonstrated commitment to build the trust and shared understandings necessary to work collaboratively under challenging political conditions'.[50] IR scholars recognise trust as a precondition for maintaining peace and securing prosperity.[51] The shared understandings help actors conduct their activities in accordance with normative expectations, and thus decrease levels of uncertainty in international relations. Using scientific knowledge stems from collaborative research practices that can reinforce shared understandings among actors.[52]

## 3        Ongoing Endeavours in US Diplomatic Negotiations

The growing number and diversity of space actors over the last decade has resulted in concerns about the possibilities of collisions and interference in outer space. The increasing uncertainty about preserving the long-term sustainability of outer space activities stimulated US diplomacy to promote existing and new norms of behaviour for conducting future responsible and sustainable space operations. The recommendations of UNISPACE III and the World Summit on Sustainable Development and UNGA Resolution on transparency and confidence-building measures in outer space activities constituted cornerstones for building the US foreign policy position during Trump's

---

presidency. The position assumed the formation of a distinct US-led international project that would include commercial partners to implement future human spaceflight programmes.[53]

Moreover, US diplomacy considered previous negotiations in plurilateral platforms. US delegates consulted their positions with the Hague International Space Resources Governance Working Group. The group dealt with the existing relevant concepts for future operations in outer space, including space resources rights.[54] Likewise, the US government and the government of Japan held international ministerial-level events on international collaboration. They promoted space for peaceful purposes and sustainability in the Tokyo Principles for International Space Exploration.[55] The participants drafted the Tokyo Principles for International Space Exploration, supporting and specifying the peaceful purposes and benefits of space exploration for humankind derived from the UN international regime on outer space.

A need to strengthen norms of behaviour through transparency and confidence-building measures stimulated US-led diplomatic negotiations about building an international coalition for implementing future human spaceflight programmes. Those drafting the Accords had previous experiences during the negotiations for the UN COPUOS Guidelines for the Long-Term Sustainability of Outer Space Activities (UN COPUOS LTS Guidelines) and other multilateral and plurilateral engagements, including the International Space Exploration Forum (ISEF) and the Hague International Space Resources Governance Working Group, and understood perspectives of partner countries and what kinds of text they were likely to be comfortable with. From the beginning the Accords were consistent with the UN COPUOS LTS Guidelines and transparency principles but were not a direct consequence of those discussions. The Accords became a distinct, US-led initiative that included commercial partners.

### 3.1    *A Need for Space Leadership during Trump's Presidency*

On the US federal executive level, President Donald Trump did not highlight space policy during his election campaign. Once elected, however, he and his space advisors preferred the US space programme to focus more on human exploration and less on researching Earth and climate sciences, with the private spaceflight industry continuing to play a significant role in space policy moving forward.[56] During his presidency, Trump amended national space policy through seven directives, releasing three orders during his first year in office.

Initially, President Trump signed Space Policy Directive 1, revising the Obama-era 2010 National Space Policy. The amended policy provided a basis for diplomatic negotiations on strengthening US national security in outer space activities. Space Policy Directive 1 also served as a starting point for promoting a sustainable space exploration programme.[57] President Trump and his advisors utilised public and private diplomatic communication strategies to create an international US-led coalition with private sector partners for a human return to the Moon, followed by missions to Mars and beyond.[58] According to Harrison and Johnson: '[I]t is also notable that Space Policy Directive 1 specifically calls for both commercial and international partners, whereas the language it replaces did not include this. While the Obama administration pursued both commercial and international partners for its exploration agenda, the explicit inclusion of these in the new space policy indicates that such partnerships were a critical component of the Trump administration's space exploration agenda'.[59] Space Policy Directive 1 directed NASA's administrator Jim Bridenstine to carry out a lunar-focused campaign.

Furthermore, the National Aeronautics and Space Administration Transition Authorization Act of 2017 encouraged the agency's diplomatic activities, and the restoration of the NSpC enabled the streamlining of national space policy and strategy development, monitoring and implementation.[60] The NSpC is a body within the Executive Office of the President of the United States, created in 1989 during the George H. W. Bush administration, which ceased operating in 1993 under the Clinton administration and was then revived in June 2017 by the Trump

---

administration.[61] The NSpC is facilitated by staff from the Executive Office of the President and headed by a civilian Executive Secretary.[62]

In his speech to the NSpC, in the autumn of 2017, Vice President Mike Pence, as the chair of the NSpC,[63] framed the need for US space leadership as a prerequisite for maintaining US national security. This point was underpinned by the development of anti-satellite technology in China and Russia, which potentially threated military effectiveness. To implement that aim, Pence underlined industry's role and declared the assembly of the NSpC UAG, a Federal Advisory Committee consisting of outside experts from industry, academia and other non-federal organisations, to support insight and expertise for the NSpC.[64] Besides that, he retasked NASA with managing its day-to-day operations 'to ensure that the interests of industry, non-Federal entities, and other persons involved in aeronautical and space activities are adequately represented by the National Space Council'.[65]

On the US national level, the NSpC UAG became the main body for providing expert input on national space policy and space exploration as its essential priority. The NSpC UAG membership is composed of representatives of industry, other non-federal entities, and other recognisable groups of persons involved in aeronautical and space activities. It also covers special government employees who may be appointed as subject matter experts to foster co-operation, and technology and information exchanges across the space industry to support US leadership in space.[66]

The NSpC UAG meets as frequently as needed, subject to the advanced approval of the NSpC UAG Designated Federal Officer but not less than once per year with the NSpC. During its second meeting, members advocated that the NSpC UAG should conduct an interagency agenda, covering US civil, military and commercial activities.[67] The agenda aimed to support the attractiveness of US space science to influence other countries and partners from the private sector to join the US-led coalition. The working agenda included promoting soft power, the economic benefits of space exploration, and international goodwill in alignment with the growing internationalisation of space programmes worldwide and the need for international collaboration from other countries, including Russia and China.[68] NSpC UAG members recognised public outreach at space-related events and their role to support domestic and international networking and encouraged its individual members to speak at public events.[69]

Following the establishment of the US Space Force and Pence's announcement to accelerate and deepen US leadership in outer space by prioritising human exploration on the Moon by 2024, members of the group proposed to form a task force with expertise at the NSpC UAG for assessing lunar exploration plans.[70] According to NASA's James Miller in a fiscal year 2022 committee detail report, the NSpC UAG was:

> actively involved in providing counsel on the US effort to safely return to the Moon and then send humans on to Mars. The NSpC UAG also produced multiple white papers that informed and advised on how the US could most effectively continue to lead international space efforts. In fiscal year 2020 there were six primary recommendations submitted to the NSpC.[71]

---

[61] The NSpC was established under the NASA Authorization Act of 1989 (Congress.gov 1988, Trump White House 2017b). The Council is composed of the following members: the Vice President, who shall be Chair of the Council, the Secretary of State, the Secretary of Defense, the Secretary of Commerce, the Secretary of Transportation, the Secretary of Homeland Security, the Director of National Intelligence, the Director of the Office of Management and Budget, the Assistant to the President for National Security Affairs, the Administrator of the National Aeronautics and Space Administration, the Director of the Office of Science and Technology Policy, the Assistant to the President for Homeland Security and Counterterrorism, the Chairman of the Joint Chiefs of Staff and the heads of other executive departments and agencies (agencies) and other senior officials within the Executive Office of the President, as determined by the Chair (Federal Register 2017).

[62] Congress.gov 1988, Trump White House 2017b.

[63] Pence 2017.

[64] Pence 2017.

[65] Pence 2017. The National Space Council Users' Advisory Group (UAG) is a Federal Advisory Committee Act (FACA) advisory committee chartered by NASA on 6 December 2017; and renewed and retasked by NASA on 4 December 2019. This committee was established pursuant to the NASA Authorization Act of 1991 (Public Law 101-611, Section 121), and Executive Order 13803, Section 6 ('Reviving the National Space Council'), signed by the President on 30 June 2017. As such, the UAG is a non-discretionary statutory Federal advisory committee and has no termination date (FACA 2017).

[66] Trump White House 2018.

[67] NSpC UAG 2018, 4.

[68] NSpC UAG 2018, 5.

[69] NSpC UAG 2018, 3–4.

[70] NSpC UAG 2019a, 3.

Those related to the Moon included:

(1) NASA should update their technology roadmap and brief the NSpC UAG in light of the Artemis and Moon and to Mars programs, provide a briefing to the NSpC UAG and, if necessary, fund a brief external review (21 October, 2019).

Three recommendations submitted in fiscal year 2019 were also partially implemented in fiscal year 2020:

(9) The NSpC UAG subcommittee recommended that the NSpC UAG form a task force to act as a 'red team' to assess the revised lunar exploration plan and develop potential alternatives. (8 April 2019)
(1) The NASA Space Technology Mission Directorate briefed its lunar technology roadmap to the NSpC UAG in January 2020.
(3) The NSpC UAG white paper, 'Assessing the Utility of a US Strategic In-Space Propellant Reserve: Economic Development in Low Earth Orbit and Cislunar Space', was delivered to the NSpC in September 2020.[72]

Additionally, Miller reported, numerous fact-finding information exchanges with NASA and the Artemis lead managers in Human Exploration and Operations Mission Directorate 'were conducted to understand the basic plans for the 2024 lunar architecture, the specific missions and visions supporting the lunar objectives, the risks associated with each element of the mission, and any broad trade spaces considered'.[73] The NSpC UAG reviewed NASA's internal independent assessment of the Artemis programme.

Discussions regarding human exploration on the Moon by 2024 continued in the NSpC UAG with the aim to secure the leading role for the US in space research. The discussions emphasised sharing activities and scientific knowledge for building transparency and confidence with international partners.[74] Towards fulfilling the aim, NSpC UAG members recommended enhancing the UN COPUOS LTS Guidelines, which could provide advantages for setting a policy framework of US commercial and civil exploration and globally inclusive space enterprises. The space policy and international engagement subcommittee of the NSpC UAG concluded there was a need to use space for improving the quality of lives instead of conducting an arms race in space.[75]

## 3.2    *Interagency Process and Space Diplomacy*

With the Space Policy Directive 1 presidential mandate and supportive recommendations coming from NSpC, individuals affiliated with the US Department of State and NASA's Office of International and Interagency Relations (OIIR) drafted the Artemis Accords. Before the Accords could be presented to international partners, the draft had to run the gauntlet through the US interagency C-175 approval process. At the time, discussions about strengthening international relations for securing US leadership in the space policy and international engagement subcommittee of the NSpC UAG in 2020.[76] During the discussions, NSpC UAG members reviewed their meetings with the Department of State, which is responsible for the C-175 process.[77] It is the standard process by which the US government authorises international negotiations.[78] C-175 gathers interagency consensus across all agencies

---

involved in such activities, and makes sure there are no concerns, objections or issues before authorisation is granted. A key issue, one state department official noted, was 'going through the interagency process and getting the authority, and then proceed with the actual negotiating process of sharing the text with other countries, getting their ideas on board, and building new versions of the text that ultimately other countries were willing to sign on to'.[79]

Partly due to rapid remote communications during the Covid-19 pandemic and mostly due to support from the NSpC and NSpC UAG, trust and existing relationships among various individuals participating in interagency negotiations, the C-175 was completed in record time.[80] Therefore, interagency trust played a crucial role in drafting and agreeing on the initial text of the Accords. Having established contacts and relationships within the US government and among federal agencies was critical to the success of the Accords.[81] What may have appeared as non-transparency from an outsider's perspective, those drafting the agreement were not allowed to talk about the Accords until the end of the C-175 process.[82] There were a number of media leaks that emerged about the Accords during the negotiations. Due to the fact that the leaks might decrease legitimacy of the process, it was challenging for the negotiators to remain silent while there were falsehoods in the media relative to the agreement.[83]

Concurrently, NSpC UAG members participated in external meetings with NASA executives and the Office of Emerging Security Challenges and the Office of Space and Advanced Technology in the US Department of State on how to communicate a demand for space resources commercialisation and norms of responsible behaviour in space as US space diplomacy priorities.[84] NSpC UAG members formally recognised the transparency and confidence-building measures in outer space activities report of the UN Secretary-General and the UN COPUOS LTS Guidelines as principal norms for conducting responsible military, civil, and commercial space activities at the international level, while also acknowledging the notoriously drawn-out negotiation process in the UN.[85] They expressed support for the Artemis Accords as an appropriate on-time solution for advancing international norms.[86] The agreement, which arose from private diplomatic communication with experts from commercial space industries, and communication with scientific professionals, emphasised various aspects of the recent outer space law regime already mentioned in existing UN Treaties and Principles.[87] Following support from the Department of State, NSpC UAG members concluded that US-led bilateral and multilateral negotiations could promote security, stability, and sustainability across the space sectors.[88]

Furthermore, one of the goals of US foreign policy, within public diplomatic communication, involved communicating norms and building international consensus towards aspects that the United States considered to be crucial in the realm of future space exploration.[89] Fulfilling space exploration objectives could get ahead of existing frameworks, so there was a need to build international norms.[90] With this in mind, the negotiators pulled together spacefaring nations to talk about principles for space exploration as an overarching framework as its plans for specific space exploration activities ramped up, especially under the Trump administration.[91] Specifically, the idea of returning astronauts to the Moon and building infrastructure there and doing that in an international context were the main rationales for the agreement.[92] Taking private diplomatic consultations with the US commercial space sector into consideration, negotiations developed the concept of the Artemis Accords as a way to start to build what might be international norms that they would want to see adopted and followed while simultaneously going forward with actual Artemis programmatic activities in the realm of space exploration.[93]

The diplomatic initiative also covered technological interoperability and a demand for sharing information about a wide range of technical systems crucial for space-based exploration, scientific discovery and the commercial utilisation of outer space.[94] Consensus on legal interpretations and technical interoperability would be crucial for rescuing astronauts and avoiding interference with activities of other actors at a location in outer space.[95] The US could persuasively use the allure of the Artemis programme to start building an international coalition of countries that would agree to a similar interpretation of those principles that then would enable lunar activities.[96]

Thus, equipped with international shared understanding and support, the US delegation, including the State Department and NASA representatives, used public diplomacy communication tools to consult, interact and influence the reception of its ideas, cultural values, and policies among various international partners. Following the optimistic opinions and recognised international trust among like-minded groups of countries, text began to emerge on how provisions of the Outer Space Treaty could be implemented in a coalition of like-minded countries.[97]

In broader context, the agreement also fitted into a comprehensive framework, as it was consistent with promoting US and Western interests and values via soft diplomacy strategies. For gaining legitimacy and confidence in diplomatic persuasion, the negotiators consulted their positions with national security bodies advocating the directions of the US Vice President. Therefore, the Artemis Accords became a complex political issue, not a specific scientific or technical problem. In order for governments to undertake their respective internal reviews, they would need to take on all the implications the US delegation had raised. There were any number of pre-existing relationships both between NASA and its counterparts and the Department of State and its counterparts internationally, which facilitated a lot of discussions and meetings. This worked exceptionally well as far as building trust concerning the issues that were raised.[98]

Supported by the adoption of US domestic legislation, the US delegation communicated via public diplomatic communication a need for the specification of a legal regime on using space resources as its national foreign policy priority was also doing in multilateral negotiations in the legal subcommittee of UN COPUOS.[99] The public diplomatic communication on the Artemis Accords derived from international activities on the Lunar Gateway in NASA's OIIR, the US State Department's Office of the Legal Adviser and Office of Space Affairs.[100] Therefore, the US diplomatic delegation refreshed its public diplomatic position for the first time since the 1970s, framing their message precisely to include that the Outer Space Treaty had not prohibited resource utilisation.[101]

The US public diplomacy effort to influence the other representatives in UN COPUOS in favour of its national space policy, including its values and ideas, and to reinforce and implement US obligations under the international law regime on space-related activities, began to gain broader international support. A diversity of views in the international community had been expressed in UN COPUOS about amending the UN Treaties and Principles on space-related activities for recent needs. Unfortunately, discussions about developing an international regime for the realisation of lunar activities with the participation of private actors were slow.[102] Therefore, US negotiators decided to communicate the Artemis Accords as a foreign policy priority to persuade the elite opinion of like-minded space agencies, because the previous bilateral and multilateral agreements and public diplomatic communication through the UN COPUOS system were less efficient. This is one reason why the Department of State and NASA chose to negotiate the initiative between like-minded space agencies with similar ideological values and to maximise the reinforcement of existing international agreements.[103]

---

[93] Interview 8 (government official, Department of State), 3 February 2022.
[94] NASA 2020, 3.
[95] US Department of State 2020.
[96] Interview 2 (space policy expert), 19 January 2022.
[97] Interview 3 (government official, Department of State), 20 January 2022; UNGA 1966.
[98] Interview 8 (government official, Department of State), 3 February 2022.
[99] UNOOSA 2017, 2; Interview 3 (government official, Department of State), 20 January 2022.
[100] Interview 3 (government official, Department of State), 20 January 2022.
[101] Interview 3 (government official, Department of State), 20 January 2022.
[102] Interview 4 (space legal expert), 24 January 2022.
[103] Interview 4 (space legal expert), 24 January 2022.

Shared understandings between NASA, private industry and political allies about building a coalition for international norms, guidelines and rules also motivated the idea of the Artemis Accords as necessary for facilitating the Artemis programme.[104] The project gained further popularity among the international community, including scientists and diplomats, because it framed exactly what many US partners had wanted to do. The Moon was the most relevant next step for space exploration and support for the Artemis Accords indicated the results of US space diplomacy to influence elites in partner countries and the possibilities for building the largest and the most diverse space coalition. Some US administration appointees perceived the Artemis programme as the last chance to lead a global coalition, due the past failures to sustain a US human space exploration programme. Its authors framed the programme and its norms of behaviour as a peaceful and prosperous way for future space exploration to persuade the international community and gain public support for the project.[105]

### 3.3  *Public Diplomacy Based on Trust*

Close professional relationships and trust within NASA and the Department of State during the negotiation process provided a strong foundation for the Artemis Accords' final support at the highest level of US federal politics.[106] The Accords created a policy and governance framework for lunar activities to operationalise a need for leadership in space policy, that stemmed from the White House during Donald Trump's presidency.[107] Within the group, an initiative for clarifying the norms of future space activities, including using space resources, provided an excellent opportunity to implement innovative space diplomacy within the existing international space law regime.

At the final phase of the negotiations, the group preferred to communicate and frame the Artemis Accord draft as a government-led product that emerged in consultation among like-minded international partners, some with previous work experience on the Lunar Gateway, to avoid misunderstandings and misinterpretations. Therefore, trust was crucial during the negotiations. The Artemis Accords were initially signed by eight countries, including the United States, Australia, Canada, Japan, Luxembourg, the United Kingdom, and the United Arab Emirates, on 13 October 2020, during a ceremony held in the first ever (due to the Covid-19 pandemic) virtual International Astronautical Congress.[108] The group of countries then built consensus around norms of behaviour, norms of operation, principles and values that countries who partnered with the US on lunar exploration would be willing to adopt in their technical operations with NASA and internationally.[109]

The Artemis Accords emerged as a public diplomacy initiative providing a convergent interpretation of relevant legal principles for enabling sustainable lunar activities. According to its final content, the Accords are a means of engaging international partners and private sector actors as a vehicle for participation in the Artemis programme and a means of implementing norms of behaviour for the conduct of future space operations.[110] The purpose of the Accords was to reinforce existing international obligations and ensure that international partners to the Artemis programme implemented those obligations, such as avoiding harmful interference, which also appears in the widely respected Outer Space Treaty. A party who wants to participate in the Artemis programme has to sign the Accords.

According to a Department of State official, the US negotiators assumed that values were integral and did not want to communicate the Artemis programme as only a technical initiative because they considered that similar values will determine the success of the co-operative effort.[111] 'The agreement was pitched at the country level, where more countries saw the benefit and the wisdom of broad principles that are espoused in the Artemis Accords and the benefit of being able to work with the US in the Artemis programme as the two go hand in hand'.[112]

At the time of concluding this research, the Accords have acquired an additional thirteen signatures from Ukraine, South Korea, New Zealand, Brazil, Poland, Mexico, Israel, Romania, Bahrain, Singapore, Colombia, France and Saudi Arabia.[113] Recently, the Biden–Harris administration declared support for the Artemis programme, including the Artemis Accords. The presidential endorsement confirmed that the programme would send another man, the first woman and person of colour to the Moon, prepare future missions to Mars and demonstrate America's and its international partners' values.[114]

---

[104] Interview 2 (space policy expert), 19 January 2022; Interview 8 (government official, Department of State), 3 February 2022.
[105] Interviews 5 (NASA OIIR official), 25 January 2022.
[106] Interview 6 (government official, National Space Council), 31 January 2022.
[107] Interview 4 (space legal expert), 24 January 2022.
[108] NASA 2019.
[109] Interview 8 (government official, Department of State), 3 February 2022.
[110] Interview 8 (government official, Department of State), 3 February 2022.
[111] Interview 8 (government official, Department of State), 3 February 2022.
[112] Interview 8 (government official, Department of State), 3 February 2022.

## 4 Conclusion

The need to sustain the peaceful use of outer space due to an ever-increasing expansion of human activity stimulates new research on space diplomacy conducted by professional diplomats and non-state actors. With the example of diplomatic negotiations for the Artemis Accords, our paper contributes to understanding diplomatic mechanisms that lead to co-operative projects in space affairs. The results of qualitative analysis of semi-structured interviews and relevant documents on US space policy have provided an in-depth exploration of the rationale for drafting the agreement, interagency authorisation and expertise facilitated by trust and the art of persuasion domestically and internationally.

 Different means of public and private diplomatic communication have also been identified in the paper. Public diplomatic communication was utilised in order to influence public opinion or other governments. The negotiators of the Artemis Accords dealt mainly with the way norms would be communicated in order to gain international support. In the context of public diplomatic communication, the importance of advanced space policy was highlighted. Private diplomatic communication aimed to enhance the attractiveness of the Artemis programme for partners from the international private sector.

 The talks mainly took place with like-minded space agencies who also had their own internal approval processes with their respective governments. Persuasion appeared mainly in the form of framing. The negotiators framed the Artemis Accords in alignment with the shared interests and values of partner countries interested in participating in the Artemis programme. Trust was present primarily during the drafting of the Accords, due to existing long-term co-operation and pre-existing relationships among the actors involved. Trust was also evident during the C-175 process.

 Our analysis of expert interviews in particular found that institutional affiliations of distinguished individuals responsible for shaping the negotiations were pivotal to initiating drafting and launching the negotiations for the Accords. The results as a whole support the theoretical assumption that space diplomacy requires not only the art of communication in the promotion of national values and influencing the preferences of other international actors, but also trust among actors involved in drafting the agreement.

 The paper also clarified the relevance of Presidential Space Policy Directives underlining US leadership and implementation of the Artemis programme. The Accords specified a framework for international co-operation for the Artemis programme and broadened the declaration of US leadership regarding norms of behaviour in space. The request from NSpC UAG had opened the door for negotiating the Accords and extensive discussions continued there on the concrete implementation of the presidential demand to strengthen US leadership. The negotiations resulted in a set of principles meant to implement the existing provisions of the Outer Space Treaty relevant for future space exploration.

 Furthermore, the analysis of diplomatic activity about the Artemis Accords revealed the complexity of space diplomacy communication mechanisms. The interviewed individuals highlighted pre-existing relationships between NASA, the Department of State and their foreign counterparts as foundational for building mutual trust and shared understandings. To make the negotiations more effective, distinguished individuals in US federal agencies did not choose UN COPUOS as the main arena for negotiations. Instead, they put emphasis on existing co-operation on security issues with government officials affiliated with partner space agencies. Yet reinforcing norms of behaviour for the participation of international actors in the Artemis programme was the main rationale for negotiating the Accords outside of the UN COPUOS process.

 Finally, the results of our research and analysis also have their limitations. Due to our focus on the domestic US process and impetus for the Accords, this analysis mainly covers the internal US interagency process and interviews were limited to US-based negotiators and legal and policy experts. Because space is a national security issue, the interviewees did not deviate from facts that were already said publicly. Furthermore, the Artemis Accords are relatively new. Therefore, there are certain limitations regarding the number of sources and prior analyses to draw from. Besides, the Artemis Accords were shaped during the Covid-19 pandemic, thus an uncommon means of diplomacy to negotiate the Accords online emerged.

---

[113] Signature countries as of 2 September 2022.
[114] Gohd 2021; The White House 2022.

There are several suggestions for further research arising from this paper and its limitations. As the paper has focused mainly on US-based negotiators and legal and policy experts, further research could focus on other signatories to the Artemis Accords. This would provide additional perspectives regarding the Artemis Accords negotiations and which type of diplomatic communication prevailed. It would certainly be appropriate to examine negotiations in space diplomacy during non-pandemic times. In this case, researchers could examine whether trust is present during a non-pandemic era, when digital communication is less present than presented in this paper. Further research on space diplomacy for the Artemis Accords could also continue with a larger author team, consisting of researchers across the Artemis partner and non-partner countries. Conducting ethnography in domestic and international institutions could also be beneficial to tracing networks of ideas and responsible individuals for space diplomacy negotiations in more detail.

**Appendix: Interviews**

| Interview 1 | space policy expert | 18 January 2022 |
|---|---|---|
| Interview 2 | space policy expert | 19 January 2022 |
| Interview 3 | government official, Department of State | 20 January 2022 |
| Interview 4 | space legal expert | 24 January 2022 |
| Interview 5 | NASA, OIIR official | 25 January 2022 |
| Interview 6 | government official, National Space Council | 31 January 2022 |
| Interview 7 | NASA, OIIR official | 1 February 2022 |
| Interview 8 | government official, Department of State | 3 February 2022 |

*Nancy Riordan*

is a Doctoral candidate in Global Governance and Human Security at the John W. McCormack Graduate School of Policy and Global Studies, University of Massachusetts, Boston, USA. Her research focuses on global space governance with particular attention to common-pool resource conflict and sustainability.

*Miloslav Machoň*

is an Assistant Professor and Research Fellow at the Department of International and Diplomatic Studies, Prague University of Economics and Business, Prague, Czech Republic. In 2020, he was appointed as a postdoctoral visiting scholar at Northeastern University, Boston, and the University of Massachusetts, Amherst, under Professor Mai'a K. D. Cross and Professor Peter M. Haas. During his studies, he collaborated closely with Professor Luboš Perek and Professor Vladimír Kopal.

*Lucia Csajková*

is a Research Fellow at the Department of International and Diplomatic Studies, Prague University of Economics and Business, Prague, Czech Republic. She won an Excellent Student Thesis Prize at the Faculty of International Relations for a seminar paper, BA and MA thesis during her studies, and completed a two-year study period in Pittsburgh, Pennsylvania. In her research, Lucia mainly focuses on international relations, sustainable development, space and environmental policy.